\documentstyle[12pt,epsfig]{article}
\begin{document}

\begin{center}
{\large
ORIENTATIONAL PHASE TRANSITION IN SOLID $C_{60}$:
BIFURCATION APPROACH
\\[2mm]}
E.E.Tareyeva and T.I.Schelkacheva
\end{center}

A simple model for the angular dependent interaction
between $C_{60}$ molecules on the face-centered
cubic lattice is proposed. The bifurcations of the
solutions of the nonlinear integral equations for
orientational distribution functions in the mean field
approximation are  analyzed and the orientational phase transition
in solid $C_{60}$ is described.
The quantitative results for the
orientational phase transition are in accordance
with experimental data.

\bigskip

1.
As is well known the mean--field approach often brings one
to the formulation of the broken space symmetry problem
in terms of the bifurcation of solutions of nonlinear
integral equations for distribution functions
or order parameters (see, e.g., the
review ~\cite{koz}).
In particular, the bifurcation approach was fruitfully
used in the case of orientational phase transitions in molecular
crystals where the  intermolecular interaction depends not only
on the distance between particles but on the molecule
orientations, too. When the temperature
decreases the free rotation of molecules fixed on the crystal
lattice sites is replaced by rotations about local frozen axes,
probably different in different sublattices.
These
orientational trasitions can be conveniently described by
changes in the $n$-particle orientational distribution functions
$f_n (\omega_1, ..., \omega_n)$. This kind of problems
was treated by use of bifurcation approach, for example,
in the papers \cite{ttdan,ttt1,rlem,shsh,sch}.  Here we will
follow our papers \cite{ttdan,ttt1,sch}.  Let us describe
briefly the main steps of the method.

In the mean--field approximation from the first
equation of the Bogoliubov hierarchy \cite{nnb}
for the equilibrium orientational
distribution functions or by minimizing the orientational free
energy one can obtain the following nonlinear integral equation
\cite{ttdan}:
\begin{equation}
g_i ( \omega_i) + {1 \over \Theta} \sum_{i \ne j} G_j
\int d \omega_i \Phi_{ij} ( \omega_i, \omega_j)
 e^{g_i (\omega_j)} = 0;
\label{main}
\end{equation}
here $\omega_i$  are the angles describing the orientation of the
molecule, e.g. Euler angles, $g_i (\omega_i) = \ln[ \frac {f_i (
\omega_i)} {G_i}]$, $f_i ( \omega_i)$ --  is the one-particle
orientational distribution function for the molecule on the site
$i$ , the constants $G_i$ are the normalization constants. If we
consider molecules fixed on the rigid lattice sites then
\begin{equation} \Phi_{ij} (\omega_i,
\omega_j) = \sum_{\nu, \tau} \Phi_{\nu, \tau} (\omega_R) u _{\nu}
(\omega_i) u_{\tau} (\omega_j), \end{equation}
where $u _{\nu} (\omega)$  --
is a full system of functions corresponding to the symmetry of
the problem.  In the case of multipole-multipole
electrostatic interaction, usually responsible for orientational
ordering in molecular crystals, we can write in the nearest
neighbor approximation
\begin{equation}
\Phi_{ij} (\omega_i, \omega_j) =
\frac{I_{l}^{2}} {R^{2l+1}} \sum_{\nu, \tau}
C_{\nu, \tau}^l (\omega_{ij})  u _{l \nu} (\omega_i) u_{l \tau}
(\omega_j),
\end{equation}
$I_l u _{l \nu}$  are the components of the molecule multipole
momentum, the coefficients  $C_{\nu, \tau} (\omega_{ij})$
are determined by the configuration  of the nearest environment,
$R$ is the nearest neighbor distance.

Assuming that the order is  different in not more than $k$
sublattices ($k\le z$), $z$ being the number of the nearest
neighbors, we will obtain from (\ref{main}) the system of $k$
nonlinear integral equations for the functions
$g_i ( \omega_i), i=1,...,k$:
\begin{equation}
\label{maink}
g_i ( \omega) + \lambda \sum_{j=1}^{z} G_j \int d \omega'
 e^{g_j (\omega')} \sum_{\nu, \tau}
C_{\nu, \tau} (\omega_{ij}) u_{l \nu} (\omega) u_{l \tau}
(\omega') = 0;
\end{equation}
$$ \lambda = {1 \over kT}
\frac{I_{l}^{2}} {R^{2l+1}}.$$

These equations differ from the well known
Hammerstein equations \cite{ham} only by the presence of the
factor $G_i$ depending  on $g_i$.

In the case of finite
domain of integration  when the fixed point principle is valid
there exists detailed theory for such equations
(see ~\cite{kra}).
We will use the standard methods (see e.g.\cite{tren}).
In the
high temperature region the system (\ref{maink}) has only
trivial solution $g_i( \omega)=0$,
corresponding to the orientationally disordered phase.
At the bifurcation points
$\lambda _{\alpha}$  new solutions with broken symmetry
appear. Real temperatures are described by $\lambda _{\alpha} >
0$.  For $\lambda = \lambda _ {\alpha} + \mu)$ the functions
$g_i^{\alpha }( \omega_i)$ can be written as series in
integer or fractional powers of
$\mu$. These powers are defined by the bifurcation
equation corresponding to the system ~(\ref{maink})
or, finally, by the fact wether the integrals
$\int d \omega u_{\nu} u_{\tau} u_{\eta}$ are zero or nonzero.

The bifurcation points are the eigenvalues of the system of
linear integral equations of the main order:
\begin{equation}
h_i ( \omega) + \lambda \sum_{j=1}^{z} G_{j}^0 \int d \omega'
 \sum_{\nu, \tau} C_{\nu, \tau} ( \omega_{ij})
u _{l \nu} (\omega) u_{l \tau} (\omega') h_j (\omega')
 = 0.
\label{lin}
\end{equation}
Here $h_i( \omega)$ are the main terms in the
expansion of $g_i(\omega)$ in powers of $\mu$.
If $ \int\! d \omega' \Phi_{ij} (\omega, \omega') \!= 0$, the
equations of the main approximation do not depend on the powers
of $\mu$ entering the expansion of $g_i(\omega)$ . It is easy to see
that
$h_i(\omega) = \sum_{\nu} h_{i}^{\nu} u_{l \nu} (\omega)$,
so  that the definite system of nonzero coefficients
$h_{i}^{\nu}$ and, consequently, the definite symmetry of the
new phase corresponds to each value $\alpha$.

Following this way we can obtain in principle
all possible broken symmetry phases and points of
transitions. In fact this program was realized in rather simple
cases of hydrogen, methane and others.

However, in the treatise of the complicated problem of the
orientational order in fullerite we simplify the program from
the very beginning. Now our aim is to describe the real
orientational order seen in the experiments, and to
obtain (without using of fitting parameters) the transition
temperature and the quantitative characteristics of the low
temperature ordered phase. Our choice of the simplified
model interaction is based on the maximum utilization of
the symmetry properties of the problem and on the
exploit of the numerical calculations
for the energy of the pair of $C_60$  molecules with
different mutual orientations, performed by other authors.

2.   First of all, let us recall that the molecule
$C_{60}$ is the result of cutting off the
vertices of icosahedron. As result of
this cutting off the 20 triangles -- the icosahedron faces --
are transformed into 20 hexagons and 12 (cutted) vertices with
5 outgoing edges -- in 12 pentagons. The edges of the resulting
body are of two types: 30 remaining parts of the initial
icosahedron edges (the $a$ type) and
$5 \times 12 = 60$ new pentagon edges (type $b$).
The detailed calculation of the
ground state of an isolated $C_{60}$ molecule shows (see, e.g.
~\cite{mol}) that the hexagons are benzol-type rings with
double bonds (and the excess of charge) placed on the
$a$-type edges, while the deficiency of charge occurs to appear
in the centers of hexagons and pentagons.

The orientational ordering in solid $C_60$ has been a subject
of extensive current experimental investigations (see, e.g.,
~\cite{braj,dav,hein,blinc} ); some theoretical
researches were performed, too
~\cite{mich,yha,lu,heid,lap,laun,gam,harr,aks}.
At ambient temperature the
molecules $C_60$ rotate almost freely with centers on the
face-centered-cubic (fcc)
lattice sites (the space group is $Fm{\bar 3}m$).

\vspace{-1.5cm}

~~~~~~~~~~~~~~~~~~~~
\begin{picture}(180,210)
\put(90,95){\vector(0,2){30}}
\put(90,95){\vector(2,0){30}}
\put(90,95){\vector(-1,-1){22}}
\put(90,95){\circle*{20}}
\put(90,135){$z$}
\put(125,95){$y$}
\put(65,65){$x$}
\put(100,100){0}
\put(60,60){\line(0,2){37}}
\put(60,113){\line(0,2){37}}
\put(60,105){\circle{16}}
\put(57,103){A}
\put(30,30){\line(0,2){37}}
\put(30,83){\line(0,2){37}}
\put(27,73){A}
\put(30,75){\circle{16}}
\put(150,60){\line(0,2){37}}
\put(150,113){\line(0,2){37}}
\put(150,105){\circle{16}}
\put(147,103){A}
\put(120,30){\line(0,2){37}}
\put(120,83){\line(0,2){37}}
\put(120,75){\circle{16}}
\put(117,73){A}

\put(60,60){\line(2,0){37}}
\put(113,60){\line(2,0){37}}
\put(105,60){\circle{16}}
\put(102,58){B}
\put(30,30){\line(2,0){37}}
\put(83,30){\line(2,0){37}}
\put(75,30){\circle{16}}
\put(72,28){B}
\put(30,120){\line(2,0){37}}
\put(83,120){\line(2,0){37}}
\put(75,120){\circle{16}}
\put(72,118){B}
\put(60,150){\line(2,0){37}}
\put(113,150){\line(2,0){37}}
\put(105,150){\circle{16}}
\put(102,148){B}

\put(150,60){\line(-1,-1){10}}
\put(130,40){\line(-1,-1){10}}
\put(135,45){\circle{16}}
\put(132,43){D}
\put(60,60){\line(-1,-1){10}}
\put(40,40){\line(-1,-1){10}}
\put(45,45){\circle{16}}
\put(42,43){D}

\put(60,150){\line(-1,-1){10}}
\put(40,130){\line(-1,-1){10}}
\put(45,135){\circle{16}}
\put(42,133){D}

\put(150,150){\line(-1,-1){10}}
\put(130,130){\line(-1,-1){10}}
\put(135,135){\circle{16}}
\put(132,133){D}
\put(0,5){Figure 1: The sublattices in fcc $C_{60}$.}
\end{picture}

\medskip

When the temperature decreases to
$T_S \approx 260 K$ the first order orientational phase
transition takes place: the sites of the initial fcc lattice
become divided between four simple cubic sublattices
(see fig.1) with own preferable molecular orientation in each
sublattice (the space group is $Pa{\bar 3}$).
Moreover, the experiments ~\cite{dav,dav1,blas}
have shown that in the
ordered state the electron--rich regions (the
interpentagon double bonds) face the electron--deficient regions
of the neighboring $C_{60}$ molecule: the centers of pentagons
or the centers of hexagons.  It was shown that the
ratio of the number of molecules in those two states is about
 $\rho_P : \rho _H = 60:40$  at the phase transition temperature
and increases when the temperature decreases. This remaining
two--level orientational disorder
is usually believed to cause the orientational glass transition
at $T_G \approx 90K$ now confirmed by various experimental
technics (see, e.g., ~\cite{glass}).

These two minima of the
intermolecular angle dependent energy were in fact obtained
through numerical calculations and were shown to be much lower
than the energies of other mutual orientations of the pair of
molecules (see, e.g., ~\cite{yha,laun}). In those calculations
the previously obtained charge distribution for the isolated
$C_{60}$ molecule ~\cite{mol} was taken into account.
 Usually recent calculations use the intermolecular
potential of Sprik et al. ~\cite{sprik}: a sum
of 6-12 and Coulomb interactions between 60 atoms $C$
and 30 double--bond centers $D$ and between each other:
\begin{eqnarray}
\Phi(C_{60}(1), C_{60}(2)) & = & \sum_{k\in C(1)} \sum_{k'\in
C(2)} 4 \epsilon \left \{ \left( \frac {\sigma_{CC}} {R_{kk'}}
\right)^{12} - \left( \frac {\sigma_{CC}}
{R_{kk'}}\right)^6\right\} \nonumber\\ & + &\sum_{k \ne k', k,
k' \in C,D} 4 \epsilon \left \{ \left( \frac {\sigma_{CD}}
{R_{kk'}} \right)^{12} - \left( \frac {\sigma_{CD}}
{R_{kk'}}\right)^6\right\} \nonumber\\  & +
&\sum_{k\in D(1)} \sum_{k'\in D(2)} 4 \epsilon \left \{ \left(
\frac {\sigma_{DD}} {R_{kk'}} \right)^{12} - \left( \frac
{\sigma_{DD}} {R_{kk'}}\right)^6\right\} + \sum_{k,k'\in C,D}
\frac {q_{k} q_{k'}} {R_{kk'}}.
\label{full}
\end{eqnarray}
Here $\epsilon =
1.293 meV, \sigma_{CC} = 3.4 \AA, \sigma_{CD} = 3.5 \AA,
\sigma_{DD} = 3.6 \AA, q_D = - 0.35e, q_C = - q_D/2.$
Rigorously speaking to write the equation
~(\ref{main}) in our case
we are interested in
the angular part of this complicated interaction represented in
terms of multipole--multipole interaction of point--like
multipoles on the sites of rigid fcc lattice with coefficients
to be calculated from ~(\ref{full}).  The general form of this
angular part is:
\begin{equation}
\Phi_{ij} (\omega_i, \omega_j) =
\sum_{l;\nu, \tau}
C_{\nu, \tau}^l (\omega_{ij})  u _{l \nu} (\omega_i) u_{l \tau}
(\omega_j)
\label{mult}
\end{equation}
with $l = 6, 10, 12, 16, 18, ...$  due to the icosahedral
molecular symmetry $I_h$ (see, e.g., ~\cite{heid}).

3. However, we will not use the multipole series {\it per se},
but will simplify the problem and develop a model orientational
interaction exploiting the symmetry properties to maximal
extend.
We will follow the main ideas of the papers ~\cite{dav,lap}
and use the restricted number of allowed orientations instead
of free continuous rotations. Let us note that a kind of more
general discrete model was proposed in Ref. ~\cite{aks}.
Let us take into account in the energy ~(\ref{mult}) only
the orientations
with pentagons, hexagons or double bonds directed
towards 12 nearest neighbors in fcc lattice.
It is important that the $C_{60}$ molecule is constructed in
such a way that if 6 of its 12 pentagons (or 6 of its 20
hexagons) face 6 nearest neighbors double bonds ($P$ and $H$
states of Lapinskas et al.~\cite{lap}),  then 6 of its 30
interpentagon double bonds (type $a$) face the remaining 6
nearest neighbors.

Now the energy matrix elements
can take only three values: $J_0$ -- the
energy of the general mutual position, $J_P$ -- pentagon
{\it versus} double bond and $J_H$ -- hexagon {\it versus}
double bond. These energies in our model
can be compared with those calculated in
~\cite{yha,lu,gam,yha1}
as functions of the angular displacements
of the molecule at (0,0,0). Following ~\cite{lap}, and putting
$J_0 = 0$ we obtain from the fig.2(b) of the paper ~\cite{yha}
$J_P = - 300K$ and $J_H = - 110K$.

The energy matrix elements $J_P$ and $J_H$ connect the
states of molecules only in the allowed orientations.
So, only allowed linear combinations of
$u_{l \nu}$ enter the energy.
The theoretical curve in ~\cite{yha} makes no difference
between the number $l$ of harmonics and describes
the effect of all of them. So,
in the framework of our model calculation it is possible
to build up the allowed functions using only the harmonics
with $l=6$: we need only their transformation properties. We
restrict ourselves to $l=6$, however the coefficients $J_P$ and
$J_H$ are not some of $C_{\nu ,\tau }^6$ given in ~(\ref{mult})
but effectively take into account higher order terms.

Let us construct the functions $P_i(\omega )$ and $H_i(\omega )$
explicitly in terms of cubic harmonics
$K_{m} \equiv K_{6,m}, m = 1, 2, ..., 13$.
The states $P_i (H_i)$ have 6 pentagons (hexagons)
and 6 double bonds directed towards 12 nearest
neighbors along different $[100]$ axes.
All functions
$P_i$ and $H_i$ are the sums of $K_m$, invariant under
the icosahedral
symmetry of the molecule (i.e. belonging to the $A_{1g}$
representation of the icosahedral group $I_h$) if
icosahedrons are naturally oriented in one
of 8 properly chosen coordinate systems.
If we choose the $z$-axis along the fifthfold icosahedral
axis, the the function
$A_1(\omega )$ can be written as ~\cite{a1g}:
 $$A_1(\omega ) =
a_0 [ Y_{6,0}(\omega ) + a_5 (Y_{6,5}(\omega ) - Y_{6,-5}(\omega
))],$$
with $a_0=-\sqrt{11}/5,$ $a_5=-\sqrt {7}/\sqrt{11},$
$Y_{lm}(\omega )$ -- spherical harmonics (taken as in
\cite{land}). After rotating the icosahedron clockwise about
$y$-axis by the angle $\phi = 58.28253$ it occurs in the
position with its three twofold axes parallel to Cartesian
axes. This is the so called standard $B$-orientation
(following ~\cite{harr}).
$P_1(\omega )$
describes the molecule rotated from the standard
orientation $B$  about $[111]$
axis through the angle $\psi = 97.76125^o$ (clockwise).
After the rotation three fifthfold icosahedron axes occur
to be directed (almost) precisely along
$[\mp 1,0,\mp
1],$ $[0,\mp 1, \mp1]$, $[\mp 1, \mp 1, 0]$, that is in the
directions of 6 nearest neighbors.
The functions $P_2 (\omega
)$,$P_3 (\omega )$ and $P_4 (\omega )$
are obtained from $P_1(\omega )$  by
subsequent counter--clockwise rotations of the molecule by
$90^o$ around $z$ axes. In analogous way one can obtain the
functions $H_i(\omega )$, with the only difference that now
$\psi'=37.76125^o$.

If written in the "standard" coordinate frame with
Cartesian axes along the cube sides these functions have the
following explicit form:
\begin{eqnarray} 
P_{1}(\omega )\!\! &\! =\! &\!\! \alpha_{P} K_{1} (\omega ) \!+ 
\!\beta_{P} [K_{8}(\omega ) +
K_{9} (\omega ) + K_{10} (\omega )] + \gamma_{P} [K_{11}(\omega
) + K_{12} (\omega ) + K_{13} (\omega )],\nonumber\\
P_{2}(\omega )\!\! & = & \!\!\alpha_{P} K_{1} (\omega )\! +\! \beta_{P} [ -\!
K_{8}(\omega ) + K_{9} (\omega ) \!-\!\! K_{10} (\omega )] +
\gamma_{P} [ -\! K_{11}(\omega ) + K_{12} (\omega ) \!-\!\! K_{13}
(\omega )], \nonumber\\
P_{3}(\omega )\!\! & = &\!\! \alpha_{P} K_{1} (\omega )\! +\! \beta _{P}
[K_{8}(\omega ) \!-\!\! K_{9} (\omega ) \!-\!\! K_{10} (\omega )] +
\gamma _{P}
[K_{11}(\omega ) \!-\!\! K_{12} (\omega ) \!-\!\! K_{13} (\omega )],
\nonumber\\
P_{4}(\omega )\!\! & = &\!\! \alpha_{P} K_{1} (\omega )\! +\! \beta _{P}
[-\! K_{8}(\omega ) \!-\!\! K_{9} (\omega ) + K_{10} (\omega )] +
\gamma _{P}
[ -\! K_{11}(\omega ) \!-\!\! K_{12} (\omega ) + K_{13} (\omega )],
\nonumber\\
H_{1}(\omega )\!\! & = &\!\! \alpha_{H} K_{1} (\omega )\! +\! \beta _{H}
[K_{8}(\omega ) + K_{9} (\omega ) + K_{10} (\omega )] +
\gamma_{H}
[K_{11}(\omega ) + K_{12} (\omega ) + K_{13} (\omega
)],\nonumber\\
H_{2}(\omega )\!\! & = &\!\! \alpha_{H} K_{1} (\omega )\! +\! \beta_{H}
[ -\! K_{8}(\omega ) + K_{9} (\omega ) \!-\!\! K_{10} (\omega )] +
\gamma_{H}
[ -\! K_{11}(\omega ) + K_{12} (\omega ) \!-\!\! K_{13} (\omega )],
\nonumber\\
H_{3}(\omega )\!\! & = &\!\! \alpha_{H} K_{1} (\omega )\! +\! \beta_{H}
[K_{8}(\omega ) \!-\!\! K_{9} (\omega ) \!-\!\! K_{10} (\omega )] +
\gamma_{H}
[K_{11}(\omega ) \!-\!\! K_{12} (\omega ) \!-\!\! K_{13} (\omega )],
\nonumber\\
H_{4}(\omega )\!\! & = &\!\! \alpha_{H} K_{1} (\omega )\! +\! \beta_{H}
[ -\! K_{8}(\omega ) \!-\!\! K_{9} (\omega ) + K_{10} (\omega )] +
\gamma_{H}
[ -\! K_{11}(\omega ) \!-\!\! K_{12} (\omega ) + K_{13} (\omega )],
\nonumber\\
~&~&~
\label{ph}
\end{eqnarray}
with $\alpha_{P}=-0.38866;$ $\beta _{P}=0.31486;$
$\gamma _{P}=-0.42877;$
$\alpha _{H}= 0.46588;$ $\beta _{H}=0.37740;$
$\gamma _{H}=0.34432$. The functions are normalized to unity.
We use the notations for cubic harmonics from
~\cite{heid} (see Appendix).

So, we suppose in the spirit of the papers
~\cite{dav,lap} that instead of free rotations only discrete
jumps between restricted number of orientations is allowed
even in the high temperature phase. Let us remind that the
number of these allowed rotations is a rather large one: the
molecule has 20 hexagons, 12 pentagons and 30 double bonds.

4. Let us write now the equations ~(\ref{lin})
for our model in the case of four sublattices of
experimentally determined $Pa{\bar 3}$
structure (see fig.1):

\begin{eqnarray}
h_1(\omega ) + \frac {\lambda} {4 \pi} \int d \omega ' [
B(\omega ,\omega ') h_2(\omega ') + A(\omega ,\omega ') h_3
(\omega ') + D(\omega ,\omega ') h_4 (\omega ')] = 0,
\nonumber\\
h_2(\omega ) + \frac {\lambda} {4 \pi} \int d \omega ' [ B(\omega ,\omega ')
h_1(\omega ') + A(\omega ,\omega ') h_4 (\omega ') +
D(\omega ,\omega ') h_3 (\omega ')] = 0,
\nonumber\\
h_3(\omega ) + \frac {\lambda} {4 \pi} \int d \omega ' [ B(\omega ,\omega ')
h_4(\omega ') + A(\omega ,\omega ') h_1 (\omega ') +
D(\omega ,\omega ') h_2 (\omega ')] = 0, \nonumber\\
h_4(\omega ) + \frac {\lambda} {4 \pi} \int d \omega ' [ B(\omega ,\omega ')
h_3(\omega ') + A(\omega ,\omega ') h_2 (\omega ') +
D(\omega ,\omega ') h_1 (\omega ')] = 0.
\label{sys1}
\end{eqnarray}

Here $\lambda=1/T$
($J_{P}$ and $J_{H}$ are taken in Kelvin)
and
$A(\omega,\omega')$, $B(\omega ,\omega ')$, $D(\omega ,\omega ')$
are the sums of interactions over
nearest neighbors of the central molecule in the sublattices
$A$, $B$ and $D$ (see fig.1), respectively.  For example, the
sum in the plain perpendicular to the $x$ axis
(the sites $D$ in fig.1) can be written
explicitly in the form:
\begin{eqnarray}
D(\omega ,\omega ') & = & \nonumber\\
=2 \{[(P_1 (\omega ) &+& P_4 (\omega )]
J_{P} + (H_1 (\omega ) + H_4 (\omega _)] J_H]\nonumber\\
 \times  [
P_2 (\omega ')& +& P_3 (\omega ') + H_2 (\omega ') + H_3 (\omega
')]\nonumber\\
+  [P_2 (\omega ) &+& P_3 (\omega ) + H_2
(\omega ) + H_3(\omega )] \nonumber\\
\times  [(P_1 (\omega ') &+&
P_4 (\omega ')) J_P + (H_1 (\omega ') + H_4 (\omega ')) J_H]
\nonumber\\
 +  [(P_2 (\omega ) &+& P_3 (\omega )) J_P + (H_2
(\omega ) + H_3 (\omega )) J_H]\nonumber\\
\times  [ P_1 (\omega ') &+& P_4 (\omega ') + H_1 (\omega ') +
H_4 (\omega ')] \nonumber\\
 +  [P_1(\omega ) &+& P_4 (\omega )
+ H_1 (\omega ) + H_4 (\omega )]\nonumber\\
 \times
[(P_2(\omega ') &+& P_3(\omega ')) J_P + (H_2 (\omega ') + H_3
(\omega ')) J_H]\}.
\label{pph}
\end{eqnarray}
The bifurcation points for the solutions of initial system are
the eigenvalues of the  system
~(\ref{sys1})
while the nontrivial solutions of
~(\ref{sys1})
define the symmetry character of the new phase.

It is easy to see that the solutions of the system ~(\ref{sys1})
can be written as linear combinations
$$h_i (\omega ) = \sum _{\nu }
h_{i}^{\nu } K_{\nu }(\omega),$$
and the system  ~(\ref{sys1}) is equivalent to the following
system of linear algebraic equations:

\begin{eqnarray}
h_1^{\nu } + \frac {\lambda} {4 \pi} \sum _{\tau=1}^{13}
[A_{\tau \nu }h_3^{\tau } + B_{\tau \nu } h_2^{\tau } + D_{\tau
\nu }h_4^{\tau }]  = 0,\nonumber\\ h_2^{\nu } + \frac {\lambda}
{4 \pi} \sum _{\tau=1 }^{13} [A_{\tau \nu }h_4^{\tau } + B_{\tau
\nu } h_1^{\tau } + D_{\tau \nu }h_3^{\tau }]  = 0,\nonumber\\
h_3^{\nu } + \frac {\lambda} {4 \pi} \sum _{\tau=1 }^{13}
[A_{\tau \nu }h_1^{\tau } + B_{\tau \nu } h_4^{\tau } + D_{\tau
\nu }h_2^{\tau }]  = 0,\nonumber\\ h_4^{\nu } + \frac {\lambda}
{4 \pi} \sum _{\tau=1 }^{13} [A_{\tau \nu }h_2^{\tau } + B_{\tau
\nu } h_3^{\tau } + D_{\tau \nu }h_1^{\tau }]  = 0,
\label{sys2}
\end{eqnarray}
$$\nu = 1, 8, 9, ..., 13.$$
Using the explicit form of the matrices
$A, B, D$  it is easy to obtain the only nonzero elements:
\begin{eqnarray} A_{1,1} & = & B_{1,1} =
D_{1,1} \equiv u, \nonumber\\ A_{8,8} & = & B_{9,9} = D_{10,10}
\equiv v, \nonumber\\ A_{11,11} & = & B_{12,12} = D_{13,13}
\equiv z, \nonumber\\ A_{8,11} & = & A_{11,8} = B_{9,12} =
B_{12,9} = D_{10,13} = D_{13,10} \equiv w. \nonumber
\end{eqnarray}

One can write the elements $u, v, z$ and $w$
in terms of the coefficients $\alpha _P$,  $\beta
_P$, $\gamma _P$, $\alpha _H$, $\beta _H$, $\gamma_H$ and energies
$J_P, J_H$ and obtain the following values: $u = 32 \cdot
5.046, v = 32 \cdot 94.127, z = 32 \cdot 7.665, w = - 32
\cdot 37.155$.
The determinant of the algebraic system ~(\ref{sys2}) is
factorized in $2 \times 2$  determinants, so that the
eigenvalues $\lambda _ \alpha $ can be easily obtained. Among
the values $\lambda _ \alpha $ there are two positive values.
The first one $\lambda_1  = 4 \pi /u$ corresponds to the solution
proportional to $K_1$ and is of no interest now.
The second
$\lambda = \lambda _ b$
is the positive solution of the equation
\begin{equation}
1 - \frac {\lambda} {4 \pi} (v + z) + \frac {{\lambda}^2}
{(4 \pi)^2} (vz - w^2) = 0,
\label{eql}
\end{equation}
namely
$ \lambda_b = 0.00364 K^{-1}$ or $ T_b = 275 K$.
The corresponding nontrivial eigenfunctions have
$h_i^1 = 0,$ and
$$ h_1^{11} - h_3^{11} = Q (h_1^8 - h_3^8) \neq 0;
 h_2^{11} - h_4^{11} = Q (h_2^8 - h_4^8) \neq 0;$$
$$ h_1^{13} - h_4^{13} = Q (h_1^{10} - h_4^{10}) \neq 0;
 h_2^{13} - h_3^{13} = Q (h_2^{10} - h_3^{10}) \neq 0;$$
$$ h_1^{12} - h_2^{12} = Q (h_1^9 - h_2^9) \neq 0;
 h_3^{12} - h_4^{12} = Q (h_3^9 - h_4^9 ) \neq 0;$$
where
$$ Q = \frac {4 \pi  - v \lambda_b} { \lambda_b w} ,$$ and
 $$ h_1^{11} + h_3^{11} = h_1^8 + h_3^8 = h_2^{11} + h_4^{11} =
h_2^8 + h_4^8 = $$ $$ h_1^{13} + h_4^{13} = h_1^{10} + h_4^{10}
= h_2^{13} + h_3^{13} = h_2^{10} + h_3^{10} = $$ $$ h_1^{12} +
 h_2^{12} = h_1^9 + h_2^9 = h_3^{12} + h_4^{12} = h_3^9 + h_4^9
 = 0.$$
If we add
the condition for the functions $h_i(\omega )$ to transform one
into another under the action of the cubic group rotation
elements which leave the fcc lattice invariant, then
only three of the coefficients remain to be independent and the
functions $h_i$ can be written in the following form:
\begin{eqnarray}
h_1(\omega ) & = & a P_1 (\omega ) + b H_1 (\omega ) + c K_1
(\omega),\nonumber\\
h_2(\omega ) & = & a P_3 (\omega ) + b H_3 (\omega ) + c K_1
(\omega),\\
h_3(\omega ) & = & a P_4 (\omega ) + b H_4 (\omega ) + c K_1
(\omega),\nonumber\\
h_4(\omega ) & = & a P_2 (\omega ) + b H_2 (\omega ) + c K_1
(\omega),\\
\label{hpk}
\end{eqnarray}
while
\begin{equation}
a \alpha _P + b \alpha _H + c = 0.
\label{fin1}
\end{equation}
Now
\begin{eqnarray}
h_1(\omega ) & = &
r(K_8(\omega ) + K_9(\omega ) + K_{10}(\omega  )) + s
(K_{11}(\omega ) +K_{12}(\omega ) +K_{13}(\omega )),\nonumber\\
h_2(\omega ) & = &
r(- K_8(\omega ) + K_9(\omega ) - K_{10}(\omega  )) + s
(- K_{11}(\omega ) + K_{12}(\omega ) - K_{13}(\omega
)),\nonumber\\
h_3(\omega ) & = & r(K_8(\omega ) - K_9(\omega )
- K_{10}(\omega  )) + s (K_{11}(\omega ) - K_{12}(\omega )
 - K_{13}(\omega )),\nonumber\\
h_4(\omega ) & = & r( - K_8(\omega ) - K_9(\omega ) +
K_{10}(\omega  )) + s ( - K_{11}(\omega ) - K_{12}(\omega )
+K_{13}(\omega )),\nonumber\\ \label{hpkk}
\end{eqnarray}
where $$r
= a \beta _P + b \beta _H, s = a \gamma _P + b \gamma _H,$$
\begin{equation}
 s = Q r.
\label{fin2}
\end{equation}

Using the numerical values given above
we obtain immediately
$ Q = - 0.3707 ,$ so that $b:a.$
We see now that the bifurcation temperature and
the ratio of the number of molecules in $P$ and $H$ states
occur to coincide with the experimental data
~\cite{dav,dav1,blas} :
\begin{equation}
\rho _P =\frac {a} {a + b} = 0.608; \qquad \rho _H=\frac {b} {a
+b} = 0.392.
\label{ro}
\end{equation}

5.
To obtain the remaining unknown coefficient we use
the equations of the second order in $\mu$.
As usually, we will calculate this coefficient from
the condition of solvability of these equations,
i.e. from the condition on right hand
sides to be orthogonal to the solutions of the homogeneous
integral equations. The second order equations have the form
(let us remind that $ g_i (\omega ) = \mu h_i (\omega ) + \mu ^2
x_i (\omega )+ ...; \lambda  = \lambda _b (1 + \mu )$):
\begin{eqnarray} x_1 (\omega ) + \frac {\lambda_b} {4 \pi} \int
d \omega ' [A(\omega , \omega ') x_3 (\omega ') + B(\omega
,\omega ') x_2 (\omega ') + D (\omega , \omega ') x_4 (\omega
')] = R_1; \nonumber\\ x_2 (\omega ) + \frac {\lambda_b} {4 \pi}
\int d \omega ' [A(\omega , \omega ') x_4 (\omega ') + B(\omega
,\omega ') x_1 (\omega ') + D (\omega , \omega ') x_3 (\omega
')] = R_2; \nonumber\\ x_3 (\omega ) + \frac {\lambda_b} {4 \pi}
\int d \omega ' [A(\omega , \omega ') x_1 (\omega ') + B(\omega
,\omega ') x_4 (\omega ') + D (\omega , \omega ') x_2 (\omega
')] = R_3; \nonumber\\ x_4 (\omega ) + \frac {\lambda_b} {4 \pi}
\int d \omega ' [A(\omega , \omega ') x_2 (\omega ') + B(\omega
,\omega ') x_3 (\omega ') + D (\omega , \omega ') x_1 (\omega
')] = R_4; \label{eqx} \end{eqnarray} where
\begin{eqnarray} R_1 &
= & h_1 (\omega ) - \frac {\lambda_b} {8 \pi} \int d \omega ' [A
(\omega ,\omega ') (h _{3})^2 + B(\omega ,\omega ') (h_{2})^2 + D
(\omega ,\omega ') (h_{4})^2]; \nonumber\\
R_2 & = & h_2 (\omega
) - \frac {\lambda_b} {8 \pi} \int d \omega ' [A (\omega ,\omega
') (h _{4})^2 + B(\omega ,\omega ') (h_{1})^2 + D (\omega
,\omega ') (h_{3})^2]; \nonumber\\
R_3 & = & h_3 (\omega ) -
\frac {\lambda_b} {8 \pi} \int d \omega ' [A (\omega ,\omega ')
(h _{1})^2 + B(\omega ,\omega ') (h_{4})^2 + D (\omega ,\omega
') (h_{2})^2]; \nonumber\\
R_4 & = & h_4 (\omega ) - \frac
{\lambda_b} {8 \pi} \int d \omega ' [A (\omega ,\omega ')
(h_{2})^2 + B(\omega ,\omega ') (h_{3})^2 + D (\omega ,\omega ')
(h_{1})^2].  \label{eqr} \end{eqnarray}

The system of nonlinear inhomogeneous equations  ~(\ref{eqx})
has nontrivial solutions for $x_i$ if the right hand sides
~(\ref{eqr}) of the system are  orthogonal to the solutions of
the homogeneous equations.  In terms of inhomogeneous linear
algebraic equations for the coefficients $x_{i}^{\mu }$
in the series for $x_i$ in spherical harmonics the analogous
condition reads that the rank of the system determinant is
equal to the rank of the determinant of the system examined
above. All 16 equations
\begin{equation}
\int d\omega R_i (\omega ) h_j (\omega )
= 0
\end{equation}
are identical and give the following equation for the
coefficients $s,r$:
 \begin{eqnarray} s^2 + r^2 + \frac
{\lambda_b } {4 \pi} \{ r^3 (M_1 v + M_3 w) + r^2 s [ 2 (M_3 v +
M_2 w) + (M_1 w + M_3 z)] + \nonumber\\ r s^2 [ 2 (M_3 w + M_2
z) + ( M_2 v + M_4 w)] + s^3 ( M_2 w + M_4 z ) \} = 0.
\label{fin3}
\end{eqnarray}
Here
$$M_1 = <8,9,10> = 0.027; M_2 = <8,12,13> = <11,9,13> =
<11,12,10> = 0.093;$$
$$M_3 = <8,9,13> = <8,12,10> = <11,9,10> = -0.012; M_4 =
<11,12,13> = -0.011$$
-- are the nonzero integrals
$$<k,m,n> = \int d\omega K_k(\omega ) K_m(\omega ) K_n(\omega).$$

Solving equations
~(\ref{fin1}), ~(\ref{fin2}) and
~(\ref{fin3}) simultaneously, we obtain finally
$ a = - 25.7;\quad b =  - 16.6;\quad c = -2.27.$
The main term in
$g_i (\omega )$ near the bifurcation point has the form:
$$g (\omega ) = - \tau [a P(\omega ) + b H(\omega ) + c K_1
(\omega )] ,$$
$$\tau = \frac {T - T_b} {T_b}.$$

6.
With this analytical solution near the bifurcation point
it is easy to find  numerical solution of the initial
nonlinear integral equations in wider region of temperatures.
The solution is of the form (~\ref{hpkk}) with the
coefficients $r(T)$ and $s(T)$ depending on the temperature and
goes in both directions from the bifurcation point  (see fig.2).
\setcounter{figure}{1}
\begin{figure}
\begin{picture}(0,170)(0,50)%
\centerline{\epsfxsize=0.7\textwidth \epsfbox{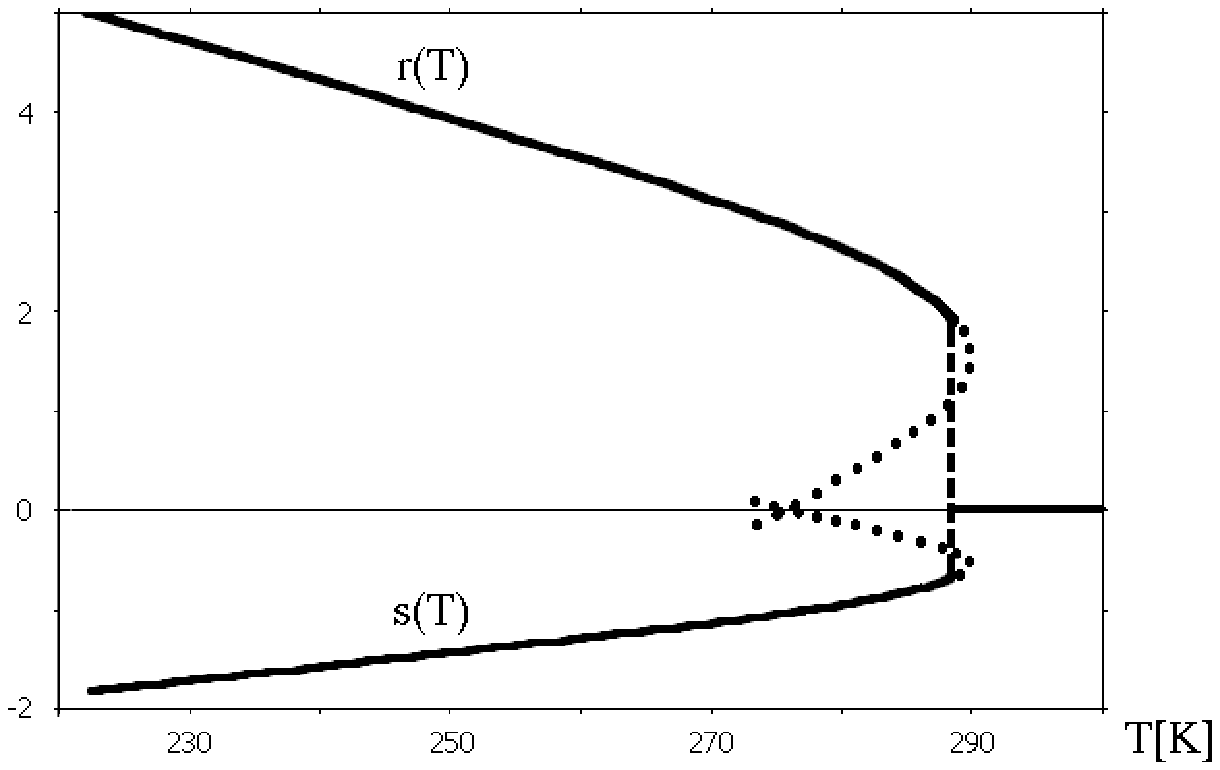}}
\end{picture}%
\caption{The coefficients $r$ and $s$ as functions of temperature.}
\end{figure}
The  solution has the turning point at $T_t = 290 K$.
The calculation of the
orientational part of the free energy of the system
$$F = \sum _{i<j} \int d \omega _i \int d \omega
_j \Phi_{i,j} (\omega _i, \omega _j) g_i(\omega _i) g_j(\omega
_j) + kT \sum _i \int d \omega _i g_i(\omega _i) \ln [4 \pi
g_i(\omega _i)]$$
was performed for the obtained solutions and the lowest
state was identified.
The solid curves in fig.2 correspond to the stable solution
and the dotted curves - to the unstable one.
At the temperature $T_c = 288.4 K$ the first
order phase transition takes place  with the jump of the
order parameter. The disordered phase  $g_i (\omega _i) = 0$  is
stable at higher temperature.  The ratio of the coefficients
$r(T)$ and $s(T)$ slightly depends on the temperature.
The relative fractions of molecules in the states $P$ and $H$ in
the ordered phase at the transition point are:
\begin{equation}
\rho _P (T_c) =\frac {a} {a + b} = 0.606; \qquad \rho _H
(T_c) =\frac {b} {a +b} = 0.394.
\label{rot}
\end{equation}
These values as well as the value of $T_c$
are in verygood agreement with the experimental data
~\cite{dav,dav1,blas}.

\medskip

7.
To conclude, we developed a simple model for angle
dependent interaction for $C_{60}$ molecules in
the fcc cubic lattice. We used rigorous analytic approach
based on the Lyapunov--Schmidt theory of bifurcation
of solutions of nonlinear integral equations
to treat this model and found the first
order orientational phase transition. The results for the
transition temperature and the characteristics of the
ordered phase are in quantitative agreement with the
experimental data.

\medskip

This work was partially supported by Russian Foundation for
Basic Researches (Grant No. 98-02-16805).

\medskip

The authors would like to thank V.N.Ryzhov and V.A.Davydov for
useful discussions.

\bigskip

{\large

Appendix.}

\bigskip

We use the expressions for cubic harmonics from ~\cite{heid}.
Let us write them explicitly.
$$K_1 = \frac {\sqrt{13}} {2 \sqrt {2 \pi}} [\frac{1}{2} (x^6 +
y^6 + z^6) - \frac{15}{4} (x^4 y^2 + x^4 z^2 + y^4 z^2 + y^4 x^2
+ z^4 x^2 + z^4 y^2) + 45 x^2 y^2 z^2],$$ $$K_8 = \frac
{\sqrt{13}} {16 \sqrt {2 \pi}} \sqrt{105} xy (1 - 18 z^2 + 33
z^4),$$ $$K_{11} = \frac {\sqrt{13}} {16 \sqrt {2 \pi}}
\sqrt{231} xy [3 (1-z^2)^2 - 16x^2y^2],$$ $$x^2 + y^2 + z^2 =
1.$$
The functions $K_9,$ $K_{10}$ and $K_{12},$ $K_{13}$
can be obtained by cyclic permutations of the coordinates from
$K_8$ and $K_{11}$, respectively.


\end{document}